\documentclass[twocolumn,prl,showpacs,superscriptaddress,amsmath,amssymb]{revtex4-1}
\usepackage{graphicx}% Include figure files
\usepackage{dcolumn}% Align table columns on decimal point
\usepackage{bm}% bold math

\begin{document}

\title{Electron-Phonon Coupling and the Soft Phonon Mode in TiSe$_2$}% Force line breaks with \\

\author{F. Weber}
\email{frank.weber@kit.edu}
\affiliation{Karlsruher Institut f\"ur Technologie, Institut f\"ur Festk\"orperphysik, P.O.B. 3640, D-76021 Karlsruhe, Germany}
\affiliation{Materials Science Division, Argonne National Laboratory, Argonne, Illinois, 60439, USA}
\author{S. Rosenkranz}
\author{J.-P. Castellan}
\author{R. Osborn}
\author{G. Karapetrov}
\affiliation{Materials Science Division, Argonne National Laboratory, Argonne, Illinois, 60439, USA}
\author{R. Hott}
\author{R. Heid}
\author{K.-P. Bohnen}
\affiliation{Karlsruher Institut f\"ur Technologie, Institut f\"ur Festk\"orperphysik, P.O.B. 3640, D-76021 Karlsruhe, Germany}
\author{A. Alatas}
\affiliation{Advanced Photon Source, Argonne National Laboratory, Argonne, Illinois, 60439, USA}

%\author{F. Weber}
%\email{fweber@anl.gov}
%\affiliation{%

%\author{F. Weber}
%\author{A. Kreyssig}%
%\affiliation{%
%\affiliation{

\date{\today}% It is always \today, today,

\begin{abstract}
  We report high-resolution inelastic x-ray measurements of the soft phonon mode in the charge-density-wave compound TiSe$_2$. We observe a complete softening of a transverse optic phonon at the \textit{L} point, i.e. $\mathbf{q} = (0.5, 0, 0.5)$, at $T \approx T_{CDW}$. Renormalized phonon energies are observed over a large wavevector range $(0.3, 0, 0.5) \le \mathbf{q} \le (0.5, 0, 0.5)$. Detailed \textit{ab-initio} calculations for the electronic and lattice dynamical properties of TiSe$_2$ are in quantitative agreement with experimental frequencies for the phonon branch involving the soft mode. The observed broad range of renormalized phonon frequencies is directly related to a broad peak in the electronic susceptibility stabilizing the charge-density-wave ordered state. Our analysis demonstrates that a conventional electron-phonon coupling mechanism can explain a structural instability and the charge-density-wave order in TiSe$_2$ although other mechanisms might further boost the transition temperature.
\end{abstract}

\pacs{71.45.Lr, 63.20.kd, 63.20.dd, 63.20.dk}% CDW, EPC, phonon measurements, phonon ab-initio calculations
\maketitle

The origin of charge-density-wave (\textit{CDW}) order, i.e., a periodic modulation of the electronic density, is a long-standing problem relevant to a number of important issues in condensed matter physics, such as the role of stripes in cuprates \cite{Kivelson03} and charge fluctuations in the colossal magnetoresistive manganites \cite{Dagotto05}. Chan and Heine derived the criterion for a stable \text{CDW} phase with a modulation wavevector $\bold{q}$ as \cite{Chan73}

\begin{equation}\label{equ1}
\frac{4\eta_q^2}{\hbar\omega_{bare}}\ge\frac{1}{\chi_q}+(2U_q-V_q)
\end{equation}

where $\eta_q$ is the electron-phonon coupling (EPC) matrix element associated with a mode at an unrenormalized energy of $\omega_{bare}$, $\chi_q$ is the dielectric response of the conduction electrons, and $U_q$ and $V_q$ are their Coulomb and exchange interactions. Static \text{CDW} order typically is taken as a result of a divergent electronic susceptibility $\chi_q$ due to nesting, i.e. parallel sheets of the Fermi surface (FS) separated by twice the Fermi wavevector $2k_f$. Electron-phonon coupling (EPC) is required to stabilize the structural distortion and, hence, an acoustic phonon mode at the \text{CDW} wavevector $\mathbf{q}_{CDW} = 2k_f$ softens to zero energy at the transition temperature $T_{CDW}$ \cite{Chan73,Gruener88}. However, when electronic probes reported only small and not well nested Fermi surfaces this scenario has been discarded for the prototypical \text{CDW} compound TiSe$_2$ \cite{Wilson77,Hughes77,Rossnagel02}.

Alternative scenarios such as indirect or band-type Jahn-Teller effects \cite{Hughes77,Rossnagel02,Whangbo92} and, most prominently, exciton formation \cite{Wilson77,Cercellier07,Rohwer11} are discussed in the theoretical as well as experimental literature. More recently, van Wezel et al. have invoked a model including exciton formation as well as EPC \cite{Wezel10a} and have shown that it can explain data from angle-resolved photoemission spectroscopy (ARPES) \cite{Wezel10b}, formerly taken as evidence of an excitonic insulating phase in TiSe$_2$ \cite{Cercellier07}. Determining the origin of \text{CDW} formation in TiSe$_2$ is all the more important with respect to the nature of superconductivity, which emerges both as function of pressure \cite{Kusmartseva09} and Cu intercalation \cite{Morosan06}. In particular, pressure induced superconductivity is expected to be closely linked to the nature of the parent \text{CDW} state.

In this letter, we report a temperature dependent study of the \text{CDW} soft phonon mode in TiSe$_2$ by high-resolution inelastic x-ray (IXS) scattering. We investigated both phonon energies as well as the phonon linewidths. We observe a complete softening of a transverse phonon branch at the \textit{L} point close to the \text{CDW} transition temperature $T_{CDW}$. Calculations of the lattice dynamical properties based on density-functional-perturbation theory (DFPT) describe our results very well and demonstrate that conventional EPC can stabilize CDW order contrary to the current discussion in the literature \cite{Rossnagel02,Cercellier07,Wezel10a}.

Our sample was a high-quality single crystal with dimensions ($3 \times 3 \times 0.05)\,\rm{mm^3}$ that shows a sharp onset of a ($2 \times 2 \times 2$) \text{CDW} superstructure at $T_{CDW} \approx 200\,\rm{K}$ as observed by synchrotron x-ray diffraction (XRD) \cite{Rosenkranz11}. The high-resolution IXS experiments were carried out at the XOR 3-ID HERIX beamline of the Advanced Photon Source, Argonne National Laboratory. The incident energy was $21.657\,\rm{keV}$ and the horizontally scattered beam was analyzed by a spherically curved silicon analyser (Reflection $[18,6,0]$). The full width at half maximum (FWHM) of the energy and wavevector space resolution was $2.4\,\rm{meV}$ and $0.07\,$\AA$^{-1}$, respectively. The components $(Q_h, Q_k, Q_l)$ of the scattering vector are expressed in reciprocal lattice units (r.l.u.) $(Q_h, Q_k, Q_l) = (h*2\pi/a, k*2\pi/a, l*2\pi/c)$; with the lattice constants $a = 3.54\,$\AA$\,$ and $c = 6.01\,$\AA$\,$ of the hexagonal unit cell $(P\overline{3}m1$, $164$).

\begin{figure}
\begin{center}
\includegraphics[width=0.95\linewidth]{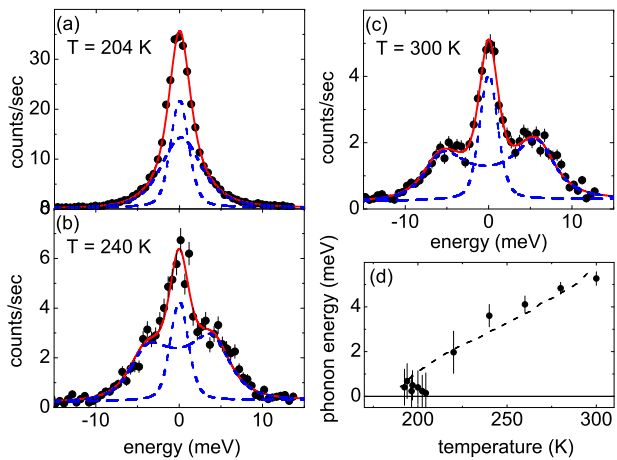}
\caption{\label{fig_1} (Color online) Temperature dependence of soft phonon mode at $\mathbf{Q} = (-0.5, 1, 0.5)$ (\textit{L} point). \textit{(a)-(c)} Energy scans for temperatures $204\,\rm{K} \le T \le 300\,\rm{K}$. Solid (red) lines are fits consisting of  a damped harmonic oscillator (inelastic) and a pseudo-voigt function (elastic) (blue dashed lines). \textit{(d)} Soft mode energy $\omega_q$ observed via IXS as function of temperature. The dashed line in \textit{(f)} is the temperature dependence deduced from thermal diffuse scattering reproduced from Ref. \onlinecite{Holt01}.}
\end{center}
\end{figure}

Measured energy spectra were fitted using a pseudo-Voigt function for the elastic line with a variable amplitude and fixed lineshape determined from scans through the \text{CDW} superlattice peak at base temperature. The phonon was fitted by a damped harmonic oscillator (DHO) function, where the energy $\omega_q$ of the damped phonon is given by $\omega_q=\sqrt{\widetilde{\omega}^{2}_{q}-\Gamma^2}$ \cite{Shirane02}, where $\widetilde{\omega}_q$ is the phonon energy renormalized only by the real part of the susceptibility, and $\Gamma$ denotes the phonon linewidth, which is closely related to the imaginary part of the susceptibility. The situation, where the phonon becomes overdamped and freezes into a periodic lattice distortion, is defined when the damping ratio $\Gamma/\widetilde{\omega}_q$ reaches its critical value, $\Gamma/\widetilde{\omega}_q=1$.

Density-functional-theory calculations were performed in the framework of the mixed basis pseudopotential method \cite{Meyer}. The exchange-correlation functional was treated in the local-density approximation (LDA). Norm-conserving pseudo-potentials for Ti and Se were constructed with Ti $3s$ and $3p$ semicore states included in the valence space. Phonon frequencies and electronic contributions to the phonon linewidth were calculated using the linear response technique or density functional perturbation theory (DFPT) \cite{Baroni01} in combination with the mixed-basis pseudopotential method \cite{Heid99}. To resolve fine features related to the Fermi surface geometry, Brillouin-zone (BZ) integrations were performed with a dense hexagonanal $24 \times 24 \times 8$ $k$-point mesh ($244$ points in the irreducible BZ). The standard smearing technique was employed with a Gaussian broadening of $\sigma = 0.15\,\rm{eV}$.  Displayed results were obtained for the undistorted hexagonal structure using experimental lattice constants and optimized internal parameters. The calculated electronic structure and lattice dynamical properties are in good agreement with previous results from LDA \cite{Jishi08, Calandra11}.
% However, note that in our calculated phonon dispersion, the leading instability occurs at the L point in agreement with experiment, whereas the preprint of Calandra et al. documents a larger negative square frequency at the M point.}

Fig.~\ref{fig_1} shows the temperature dependence of the soft phonon mode measured at the \textit{L}-point, i.e. $\mathbf{Q} = (-0.5, 1,0.5)$, from temperatures just above $T_{CDW}$ up to room temperature (Fig.~\ref{fig_1}a-c). At room temperature we find a phonon energy $\omega_q = 5.3\,\rm{meV}$ in good agreement with previous neutron scattering data \cite{Wakabayashi78}. On lowering the temperature, $\omega_q$ softens to zero energy (within the experimental error) already at $T = 204\,\rm{K}$. At $T_{CDW} \le T \le 204\,\rm{K}$, the phonon is overdamped and can only be observed as a broad quasi-elastic peak underneath the sharp elastic line, which becomes gradually more dominant below $200\,\rm{K}$ (not shown). For $T \le 190\,\rm{K}$ low energy phonon scattering directly at the \textit{L} point was undetectable due to the strong \text{CDW} superstructure peak. Overall, our results roughly agree with energies of the soft mode deduced from thermal diffuse scattering \cite{Holt01}.

\begin{figure}
\begin{center}
\includegraphics[width=0.95\linewidth]{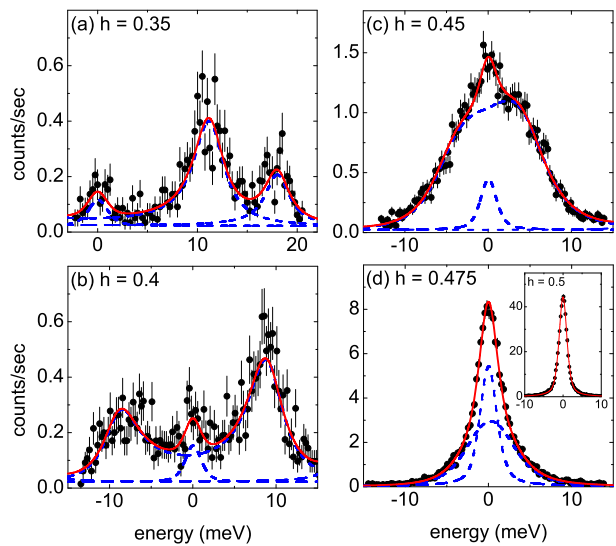}
\caption{\label{fig_2} (Color online) Energy scans at $\mathbf{Q} = (-h, 1, 0.5)$, $h = 0.35 - 0.5$, for temperature $T = 190\,\rm{K}$ ($\approx T_{CDW}$). Solid lines represent the total fit result consisting of a damped harmonic oscillator for the inelastic and a pseudo-voigt function for the elastic scattering (blue dashed lines). The inset of \textit{(d)} shows the scan at $h = 0.5$, which can be well described just by the pseudo-voigt function as used in \textit{(a)-(d)}.}
\end{center}
\end{figure}

\begin{figure}
\includegraphics[width=0.95\linewidth]{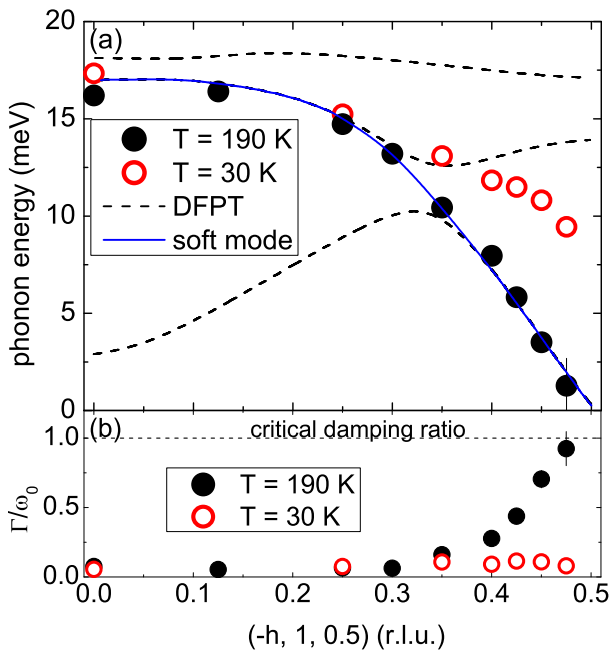}
\caption{\label{fig_3} (Color online) \textit{(a)} Dispersion of the soft phonon mode at $T = 190\,\rm{K}$ ($\approx T_{CDW}$, filled symbols) and $T = 30\,\rm{K}$ (open circles) along $\mathbf{Q} = (-h, 1, 0.5)$, $h = 0 - 0.475$. Dashed lines are results of \textit{DFPT} calculations using the experimental crystal structure and $\sigma = 0.15\,\rm{eV}$. The solid (blue) line indicates the dispersion of the soft mode character across the anti-crossing at $h \approx 0.33$. Frequencies of the upward dispersing branch (e.g. see fig. 2(a)) agree well with the calculation but are not shown for the sake of clarity. \textit{(b)} Damping ratio of the \textit{DHO} at $T = 190\,\rm{K}$ (filled symbols) and $T = 30\,\rm{K}$ (open circles).}
\end{figure}

We studied in detail the dispersion of the transverse optic (TO) phonon starting at $17\,\rm{meV}$ at \textit{A} and ending in the \text{CDW} related soft mode at \textit{L}, i.e. $\mathbf{q} = (-h, 0, 0.5)$ and $0 \le h \le 0.475$ \footnote{Phonon scattering at $h = 0.5$ was undetectable for $T \le 190\,\rm{K}$ due to strong elastic scattering.}.  We note that at $h \approx 0.33$ there is an anti-crossing of this TO branch with an acoustic-like upward dispersing branch. The character of the soft mode is transferred to the energetically lower branch for $h > 0.33$. For simplicity, we speak of the soft-mode dispersion as represented by the dashed (blue) line in Fig.~\ref{fig_3}a, which involves parts of the TO branch ($h < 0.33$) and the acoustic-like branch ($h > 0.33$). A set of energy scans taken at $T = 190\,\rm{K}$ is shown in Fig.~\ref{fig_2}.

At $h = 0.35$ (Fig.~\ref{fig_2}a), we observe two phonon peaks, where the one at lower energies corresponds to the character of the soft mode starting at $17\,\rm{meV}$ at \textit{A}. The second peak near $17\,\rm{meV}$ represents the continuation of the upward dispersing acoustic-like branch in good agreement with structure factor calculations by DFPT. This mode does not show any significant temperature dependence. Results of its dispersion are not shown in Fig.~\ref{fig_3} for the sake of clarity and will be discussed in detail elsewhere.

The energy of the phonon peak related to the soft mode already decreases for $h \ge 0.25$ and, after the anti-crossing, shows a quasi-linear dispersion relation for $h \ge 0.4$ (Fig.~\ref{fig_2}b-d), which extrapolates to a zero phonon energy at $h = 0.5$ (Fig.~\ref{fig_3}a). At $h=0.5$ however, the strong \text{CDW} superlattice reflection made it impossible to directly measure the quasielastic scattering expected for the critically damped phonon. The calculated dispersions agree very well with these results. Corresponding to the strong renormalization of the phonon energy, we observe an increase of the damping ratio $\Gamma/\widetilde{\omega}_q$, which stays just below $1$ for $h = 0.475$. But just like the energy, the damping ratio also shows linear behavior for $h \ge 0.4$ and extrapolates to an overdamped phonon at $h = 0.5$ (Fig.~\ref{fig_3}b).

The range in \textbf{q} over which phonon energies are renormalized as function of temperature allows us to infer how sharp corresponding electronic signatures, e.g. in $\chi_q$ in a nesting scenario, might be \cite{Gruener88}. We note that, in principle, the \textbf{q} dependence of the EPC matrix element $\eta_q$ can determine the periodicity of the low temperature superstructure as well \cite{Weber11}. Because the soft mode at $T = 300\,\rm{K}$ is still far from being unrenormalized ($\Gamma/\widetilde{\omega}_q (T = 300\,\rm{K}) = 0.5$), we compare the dispersion at $T = 190\,\rm{K}$ with results for $T = 30\,\rm{K}$ in the \text{CDW} ordered state, as the latter give a good idea of the range of renormalized phonons. Fig.~\ref{fig_3} clearly shows, that the phonon energies over the range $0.3 \le h \le 0.5$ are renormalized across the \text{CDW} transition. This large range in $h$ is in contrast to compounds, where a sharp FS nesting drives the \text{CDW} formation and renormalized phonon energies are observed only over a much smaller \textbf{q} range \cite{Renker73,Hoesch09}.

\begin{figure}
  \begin{center}
    \includegraphics[width=0.95\linewidth]{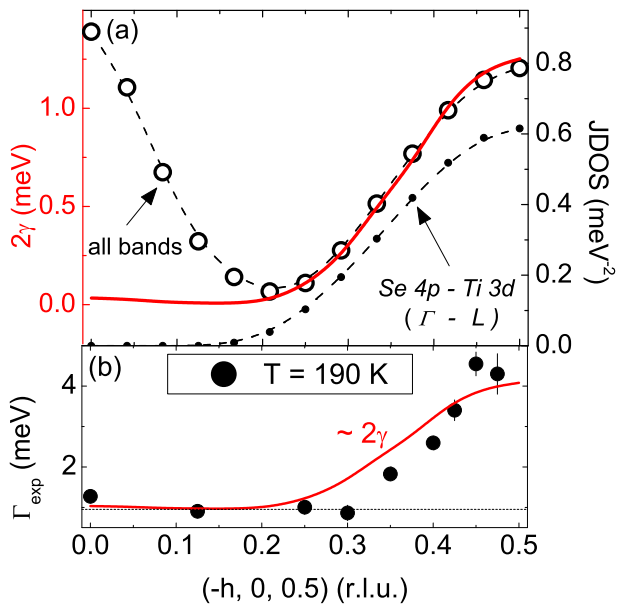}
    \caption{\label{fig_4} (Color online) \textit{(a)} Results of the calculation are shown as function of $h$ along the $A - L$ line, i.e. $\mathbf{q} = (h, 0, 0.5)$ with $h = 0 - 0.5$: (1) the calculated electronic contribution to the phonon linewidth $2\gamma$ of the soft phonon mode (FWHM, solid line, left hand scale), (2) the total electronic \textit{JDOS} (open symbols, right-hand scale) and (3) interband contributions to the \textit{JDOS} between the Se $4p$ and Ti $3d$ bands near $\Gamma$ and $L$  (filled symbols, right-hand scale). \textit{(b)} Experimental linewidth $\Gamma$ (dots) compared to scaled $2\gamma$ ($\times2.5$+offset, solid line).}
  \end{center}
\end{figure}

Comparing the experimental results in more detail to the calculations, it is important to note that DFPT is a single particle response theory and, therefore, cannot include, e.g., exciton formation. However, EPC is calculated in great detail and, in particular, the electronic contribution to the phonon linewidth $2\gamma$ can be deduced (Fig.~\ref{fig_4}). $2\gamma$ of a specific phonon mode in turn can be approximately expressed as \cite{Heid02}
\begin{equation}\label{equ2}
2\gamma=4\pi\omega_{bare}\times|\eta_q|^2\times JDOS
\end{equation}
where the electronic joint density-of-states (\textit{JDOS}) in turn is closely related to the imaginary part of the electronic susceptibility $\chi_q$. We see that the \textbf{q} range over which $2\gamma$ shows a strong enhancement matches closely the range of observed renormalized phonon energies (Fig.~\ref{fig_3}a). Furthermore, $2\gamma$ also qualitatively explains the \textbf{q} dependence of the observed linewidth $\Gamma$ very well (see Fig.~\ref{fig_4}b). Quantitatively, the calculated linewidth is a factor of $2.5$ smaller than observed in experiment. Here, we note that similar observations have been made in the CDW compound NbSe$_2$ \cite{Weber11} and the conventional superconductor YNi$_2$B$_2$C \cite{Pintschovius08}. In the latter, it is evident that only the lowest acoustic modes, which exhibit strong EPC, show this effect of too small calculated linewidths, whereas the linewidths of higher energy modes are predicted in much better agreement \cite{Weber11a}. Therefore, we think that soft modes as in TiSe$_2$ are subject to an increased anharmonicity close to the structural phase transition reflected in an increased linewidth, which is not captured in our theory.

We now discuss the different contributions to $2\gamma$, i.e. $|\eta_q|$ and the $JDOS$, according to equ.~\ref{equ2}. The fact that $2\gamma$ for the soft phonon does not follow the increase of the total \textit{JDOS} below $h = 0.2$ is due to the presence of different bands near the Fermi energy. Whereas the \textit{JDOS} at small $h$ is dominated by scattering processes within the Se $4p$ bands around the $\Gamma$ point, at large $h$ scattering between Se $4p$ and Ti $3d$ bands dominate (Fig.~\ref{fig_4}a). Apparently, $\eta_q$ of the soft mode only samples the latter scattering processes. Using equation ~\ref{equ2}, $\omega_{bare}=17\,\rm{meV}$ and the partial \textit{JDOS} of Se $4p$ - Ti $3d$ scattering processes, we can calculate $|\eta_q|$. However, this is only possible for $h\ge 0.25$ as scattering processes of the same type seem to contribute to the EPC of different phonon modes for $h<0.25$ and a clear determination of the relevant \textit{JDOS} is not possible anymore. We calculate $|\eta_q|=0.098\,\rm{meV}$ at $h = 0.5$ and a $33\%$ decrease towards $h=0.25$. For the same \textbf{q} range, the partial \textit{JDOS} of Se $4p$ - Ti $3d$ scattering processes decreases by a factor of six. Hence, the $\mathbf{q}$ dependence of the $JDOS$ is the dominant factor in the large increase of the phonon linewidth. The corresponding phonon renormalization then leads to the structural instability at the $L$ point.

Our results explain the \text{CDW} phase transition in TiSe$_2$ in the conventional picture of an increased electronic susceptibility at $\textbf{q}_{CDW}$ aided by a strong EPC matrix element. This is not in contradiction to ARPES data, on the basis of which FS nesting previously has been ruled out as a driving force of the \text{CDW} transition in TiSe$_2$ \cite{Rossnagel02}. Indeed, our results confirm that TiSe$_2$ does not exhibit a divergent $JDOS$. However, the peak in the $JDOS$ is connected with an enhancement of $Re\,\chi$ \cite{Calandra11}, which can, according to equ.~\ref{equ1}, tip the balance between undistorted and \text{CDW} ordered phases if stabilized by a sufficiently large $\eta_q$. We note that DFPT is a zero temperature approach and, therefore, cannot calculate a transition temperature corresponding to, e.g., the calculated EPC strength. So, EPC might be only one ingredient for driving the \text{CDW} transition in TiSe$_2$, but our analysis shows that its importance has been greatly underestimated.%. Indeed, time-resolved ARPES results measured at $T = 100\,\rm{K}$ show timescales typical for an electronic mechanism \cite{Rohwer11}. Effects of EPC might be present only at lower temperatures. Still, it cannot be neglected in particular with respect to the pressure induced superconductivity in TiSe$_2$ with a maximum $T_c \approx 2\,\rm{K}$ \cite{Kusmartseva09}.

In conclusion we have reported a detailed analysis of the soft phonon behaviour in the \text{CDW} compound TiSe$_2$. The experimental observations are well reproduced by ab-initio calculations, which demonstrate that electron-phonon-coupling is strong enough to stabilize a structural phase transition in TiSe$_2$ at finite temperatures. Our work emphasizes the essential part of electron-phonon coupling matrix elements and provides a new route to seemingly mysterious phase transitions as far as the electronic footprint is concerned.

\begin{acknowledgments}
We acknowledge valuable discussions with Jasper van Wezel and Mike Norman. Work at Argonne was supported by U.S. Department of Energy, Office of Science, Office of Basic Energy Sciences, under contract No. DE-AC02-06CH11357.
\end{acknowledgments}

%\bibliographystyle{apsrev.bst}
%\bibliography{bibtex_endnote}

\end{document}